\documentclass[aps, prb, twocolumn, showpacs, superscriptaddress]{revtex4-1}

\usepackage{amsmath,amssymb}
\usepackage{graphicx}
\usepackage{epstopdf}
\usepackage{dcolumn}
\usepackage{bm}
\usepackage{ulem}

\usepackage{color}

\begin{document}

\title {Kondo spin liquid in Kondo necklace model: Classical disordered phase versus symmetry-protected topological state}
\author{Yin Zhong}
\email{zhongy05@hotmail.com}
\affiliation{Center for Interdisciplinary Studies $\&$ Key Laboratory for
Magnetism and Magnetic Materials of the MoE, Lanzhou University, Lanzhou 730000, China}
\author{Yu-Feng Wang}
\affiliation{Center for Interdisciplinary Studies $\&$ Key Laboratory for
Magnetism and Magnetic Materials of the MoE, Lanzhou University, Lanzhou 730000, China}
\author{Han-Tao Lu}
\email{luht@lzu.edu.cn}
\affiliation{Center for Interdisciplinary Studies $\&$ Key Laboratory for
Magnetism and Magnetic Materials of the MoE, Lanzhou University, Lanzhou 730000, China}
\author{Hong-Gang Luo}
\email{luohg@lzu.edu.cn}
\affiliation{Center for Interdisciplinary Studies $\&$ Key Laboratory for
Magnetism and Magnetic Materials of the MoE, Lanzhou University, Lanzhou 730000, China}
\affiliation{Beijing Computational Science Research Center, Beijing 100084, China}

\date{\today}

\begin{abstract}
  { We study possible topological features of Kondo spin liquid
    phase in terms of the one- and two-dimensional Kondo necklace
    models within the frame work of quantum O(N) non-liner sigma model
    (NLSM). In the one-dimensional case, it is found that the bulk
    properties of the Kodno spin liquid phase are similar to the
    well-known Haldane phase at strong coupling fixed point. The
    difference between them mainly comes from their boundaries due to
    the effect of the topological term. In the two-dimensional case,
    the system can be mapped onto an O(4)-like NLSM with some O(3)
    anisotropy. Interestingly, we find that if hedgehog-like point
    defects are included together with the restoration of the full
    O(4) symmetry, our model is identical to a kind of SU(2)
    symmetry-protected topological (SPT) state. Additionally, if the
    system has the O(5) symmetry instead, the effective NLSM with
    Wess-Zumino-Witten term is just a description of the surface modes
    of a three-dimensional SPT state, though such O(5) NLSM could not
    be a proper description of Kondo spin liquid phase due to its
    gaplessness. We expect that the discussions might provide useful
    threads to identify certain microscopic bilayer antiferromagnet
    models (and related materials), which can support the desirable
    SPT states.}
\end{abstract}

\maketitle

\section{Introduction} \label{intr}

It is still challenging to understand the emergent quantum phases and
corresponding quantum criticality in heavy fermion
systems.\cite{Doniach,Sachdev2011,Rosch,Vojta,Custers1,Custers2,Matsumoto,Senthil2003,Senthil2004,Pepin2005,Kim2010,Senthil2010,Zhong2012e}
To tackle the problem, the Kondo lattice model has been introduced,
which is believed to capture the nature of interplay between Kondo
screening and the magnetic interaction, namely, the
Ruderman-Kittel-Kasuya-Yosida (RKKY) exchange interaction, mediated by
conduction electrons among localized spins.\cite{Tsunetsugu} The
former effect favors a nonmagnetic spin singlet state in strong
coupling limit while the latter tends to stabilize usual magnetic
ordered states in weak coupling
limit.\cite{Lacroix,Zhang2000,Capponi,Watanabe,Zhang2010,Zhang2011}

However, if we are only interested in the half-filled system with
low-lying spin excitations, a further simplified theoretical model,
namely, Kondo necklace model originally introduced by Doniach, will be
an alternative starting point.\cite{Doniach} The Kondo necklace model
is widely studied in many analytical and numerical methods, and most
of them focus on the one-dimensional (1D)
case.\cite{Scalettar,Moukouri,Nishino,Zhang2000k,Yamamoto2003,Continentino2004,Brenig2006,Langari2006,Continentino2007,Reyes2007,Langari2009,Valencia2010a,Valencia2010b,Motome2010}
It is now believed that in 1D, the system is always gapped and the
ground-state is in the Kondo spin singlet phase (also named Kondo spin
liquid phase). The issues for higher dimension are still open.

Recently, topological properties of matter which are beyond the
classic Landau symmetry-breaking theory have aroused great
interest.\cite{Wen2004} Particularly, the idea of the
symmetry-protected topological (SPT) state is proposed to classify
many possible distinct insulating phases. The SPT states are
bulk-gapped quantum phases with symmetries, and have gapless or
degenerate boundary states as long as the symmetries are not
broken.\cite{Chen2011a,Chen2013,Levin2012,Lu2012,Senthil2013,Vishwanath}
A simple example of the SPT state is the well-known Haldane phase in
$S=1$ antiferromagnetic spin chain, which is protected by its SO(3) spin
rotation symmetry. Free fermion topological insulators are also SPT
phases protected by their time reversal symmetry and U(1) charge
conservation
symmetry.\cite{Bernevig2006,Kane2006,Wiedmann,Shen2009,Hasan2010,Qi2011,Kitaev2008,Schnyder2008}

We note that the Haldane phase in $S=1$ can be well described by two
coupled antiferromagnetic $S=1/2$ spin chains and the corresponding
non-linear sigma model description correctly captures the non-trivial
topological
feature.\cite{Sachdev2011,Haldane1983,Jolicoeur1994,Gu2009,Pollmann2012}
It is noted that the Kondo necklace model may also be considered
effectively as two coupled antiferromagnetic $S=1/2$ chains (two-leg
ladders) but with antiferromagnetic interaction between two
chains. Since the Haldane phase is topologically nontrivial, it is
interesting to study the Kondo spin liquid phase in the
one-dimensional Kondo necklace model (also its higher-dimensional
extensions) and to see whether it has hidden topological feature or
not.

To uncover the possible topological properties, we will utilize the
effective quantum O(N) non-liner sigma model (NLSM) field theory
supported by topological $\theta$ terms.\cite{Sachdev2011,Haldane1983}
It is found that at strong coupling fixed point, the Kodno spin liquid
phase and the celebrated Haldane phase have similar bulk properties
described by NLSM without topological term. The main
distinction of these two phases comes from their boundary: on either
end of the open chains, the Haldane phase has fractionalized $S=1/2$
spins while the Kodno spin liquid does not support such degree of
freedom. Therefore, we may consider the Kondo spin liquid in Kondo
necklace model as a classical disordered phase.

We next analyze the topological properties of Kondo spin liquids in
high dimensional Kondo necklace models. It seems that if hedgehog-like
topological defect is ignored, the NLSM tells us that those Kondo spin
liquids are still topologically trivial unless spin-orbit
coupling-like elements are introduced. On the other hand, and more
interestingly, when hedgehogs-like point defects are included and the
symmetry is expanded to O(4) or SU(2), our model is identical to a
kind of SU(2) SPT models. This indicates that certain microscopic
bilayer antiferromagnet models may to be found to support the
desirable SPT state. Furthermore, if the symmetry is expanded to O(5),
the effective action just describes the surface state of a
three-dimensional (3D) SPT state, though such O(5) NLSM could not be a
proper description of Kondo spin liquid phase due to its gaplessness.

The remainder of the paper is organized as follows. In Sec.\ref{sec1},
two coupled spin-1/2 antiferromagnetic Heisenberg chains and the
corresponding NLSM are introduced. Three limit cases are analyzed in
the next section, which correspond to the decoupled case, Haldane
phase, and featureless disordered state, respectively. In
Sec.\ref{sec3}, the Kondo necklace model is extended and studied in
detail. In Sec.\ref{sec4}, the discussion is generalized to higher
dimension by employing an O(4) and O(5) NLSM with possible topological
terms, and the issue of topological Kondo insulator is briefly
discussed. Finally, a concise conclusion is devoted to Sec.\ref{sec5}.

\section{Two coupled spin-1/2 antiferromagnetic Heisenberg
  chains}\label{sec1}

First, we consider the two coupled spin-1/2 antiferromagnetic
Heisenberg chains\cite{Gogolin}
\begin{equation}
H=J_{1}\sum_{i}\vec{S}_{1i}\cdot\vec{S}_{1i+1}+J_{2}\sum_{i}\vec{S}_{2i}\cdot\vec{S}_{2i+1}+V\sum_{i}\vec{S}_{1i}\cdot\vec{S}_{2i},
\end{equation}
where $J_{1}$, $J_{2}$ denotes the antiferromagnetic ($J_{1},J_{2}>0$)
Heisenberg exchange coupling for each chain, respectively. The last
$V$-term describes the coupling between spins on different chains. The
sign of $V$ is crucial and the point will be elaborated later. Noting
that the four-spin ring exchange interaction is not considered in the
present model.\cite{Gogolin,Capponi2013} This interaction, if
included, is able to lead to staggered or uniform dimerization
phase between two chains and drives the two chains to the
SU(2)$_{k=2}$ gapless fixed points.

After standard treatment,\cite{Sachdev2011,Haldane1983} the low energy
properties of the above coupled antiferromagnetic Heisenberg chains
can be described by the following non-liner sigma model (NLSM) with
appropriate topological terms as
\begin{eqnarray}
&&S=\sum_{i=1,2}\int d\tau dx\mathcal{L}_{i}+i2\pi SQ_{i}+\int d\tau dx V(\vec{n}_{1}\cdot\vec{n}_{2}),\nonumber\\
&&\mathcal{L}_{i}=\frac{1}{2c_{i}g_{i}}(\partial_{\mu}\vec{n}_{i})^{2}=\frac{1}{2c_{i}g_{i}}[(\partial_{\tau}\vec{n}_{i})^{2}+c_{i}^{2}(\partial_{x}\vec{n}_{i})^{2}],\nonumber\\
&&Q_{i}=\frac{1}{4\pi}\int d\tau dx \vec{n}_{i}\cdot(\partial_{\tau}\vec{n}_{i}\times\partial_{x}\vec{n}_{i}).\label{eq1}
\end{eqnarray}
Here, we have defined $g_{i}\sim 1/S$ and $c_{i}\sim J_{i}S$ with
$S=1/2$ and $\vec{n}_{i}\sim(-)^{i}\vec{S}_{i}/S$ is a three-component
vector satisfying
$\vec{n}_{i}^{2}=n_{ix}^{2}+n_{iy}^{2}+n_{iz}^{2}=1$. The topological
term $i2\pi SQ_{i}$ ($Q_{i}=0,1,2...$) represents the fundamental
quantum features, which are absent in classical spin
systems.\cite{Sachdev2011,Haldane1983} The topological invariant
$Q_{i}$ has a simple physical interpretation: if one regards the field
configuration $\vec{n}_{i}=\vec{n}_{i}(\tau,x)$ as a map from the
two-dimensional (2D) space-time with periodic boundary condition to
the surface of a unit sphere, the topological invariant is simply the
number of times of the unit sphere being wrapped by this
projection.\cite{Sachdev2011} For spin-half-integer like $S=1/2$
systems, the topological term contributes to the partition function
like $(-1)^{Q}$, while for the spin-integer (e.g., $S=1$) situation,
such topological term does not affect the partition function.

It is well-known that, the ground state and low-lying excitations of
the spin-1/2 (SU(2)) antiferromagnetic Heisenberg chain can be
described as a gapless Luttinger liquid with some marginal logarithmic
corrections\cite{Giamarchi}, which is given by the Hamiltonian
\begin{eqnarray}
H=\frac{1}{2\pi}\int dx\left[u K(\partial_{x}\theta)^{2}+\frac{u}{K}(\partial_{x}\phi)^{2}\right],\label{eq2}
\end{eqnarray}
where $u$ denotes the non-universal velocity and $K=1/2$ is the
Luttinger parameter. The commutation relation of $\phi(x)$ and
$\theta(x)$ fields reads
$[\phi(x),\partial_{y}\theta(y)]=i\pi\delta(x-y)$. The right or left
fermion operator is defined by $\psi_{\sigma=R,L}(x)\propto
e^{-i(\phi(x)-\sigma\theta(x))}$. With these abelian bosonization
formula in hand, the spin-spin correlation is readily found as
$\langle\vec{S}(x,0)\cdot\vec{S}(0,0)\rangle\simeq
C_{1}+C_{2}(-1)^{x}\frac{\log^{1/2}(x)}{x}$ while the dimer-dimer
correlation $\langle\phi_{0}(x)\phi_{0}(0)\rangle$ [where
$\phi_{0}(x)\sim(-1)^{x}\vec{S}(x)\cdot\vec{S}(x+1)$, denoting the
dimer (valence bond) order parameter] has the same power-law-like
decay. [Note that the alternative and more sophisticated
$SU(2)_{k=2S}$ Wess-Zumino-Witten (WZW) model gives nice description
of the gapless intermediate coupling fixed point for all half-integer
antiferromagnetic chains.\cite{Affleck1986,Affleck1987} One may also
notice that in Sec.\ref{sec4}, the $SU(2)_{k=1}$ WZW model is used to
describe the edge states of some SU(2) SPT states.]

Importantly, it is widely believed that the NLSM with the topological
term $i\pi Q$ provides the same low-energy description of the spin-1/2
antiferromagnetic Heisenberg chain as well, while the NLSM combined
with $i2\pi Q$ leads to a correct description for the spin-1
Heisenberg chain.\cite{Sachdev2011,Haldane1983}

\section{Three limit situations} \label{sec2}

\begin{figure}
\includegraphics[width=0.8\columnwidth]{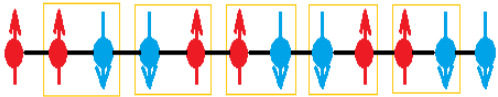}
\caption{\label{fig:1} A cartoon for the edge state of the Haldane
  phase in $S=1$ antiferromagnetic chain. One $S=1$ spin is
  fractionalized into two $S=1/2$ objects. Those $S=1/2$ spins from
  neighboring sites form singlet states except for the ones on the
  boundary.}
\end{figure}

In this section, let us inspect several simple and useful limits to
get some insights. First, the simplest case is $V=0$, where two
spin-1/2 antiferromagnetic chains are decoupled and each of them is
described by a gapless Luttinger liquid like Eq.(\ref{eq2})
separately. Next we consider $V\rightarrow-\infty$ (ferromagnetic
coupling between two chains) which fixes
$\vec{n}_{1}=\vec{n}_{2}\equiv\vec{n}$, and in this case,
Eq.(\ref{eq1}) becomes
\begin{eqnarray}
&&S=\int d\tau dx\frac{1}{2cg}(\partial_{\mu}\vec{n})^{2}+i2\pi 2SQ,\nonumber\\
&&Q=\frac{1}{4\pi}\int d\tau dx \vec{n}\cdot(\partial_{\tau}\vec{n}\times\partial_{x}\vec{n}).\label{eq3}
\end{eqnarray}
Obviously, this action just represents a spin-2S antiferromagnetic
chain with the topological term $i2\pi 2SQ$. The coupling strength and
velocity are renormalized as
$\frac{1}{cg}=\frac{1}{c_{1}g_{1}}+\frac{1}{c_{2}g_{2}}$,
$c=\sqrt{c_{1}c_{2}}\sqrt{\frac{c_{1}g_{2}+c_{2}g_{1}}{c_{1}g_{1}+c_{2}g_{2}}}$.
We can see that when $V$ is large negative, the coupled spin-1/2
chains are effectively described by a single spin-1 antiferromagnetic
chain. And it is well established that the spin-1 antiferromagnetic
Heisenberg chain always situates in its gaped phase (Haldane phase),
and there exist two free spin-1/2 spins on the boundary if the open
boundary condition is imposed (as shown in
Fig. \ref{fig:1}).\cite{Sachdev2011,Haldane1983,Jolicoeur1994,Gu2009,Pollmann2012}
In other respect, the Haldane phase is a 1D
symmetry-protected-topological phase since as long as the SO(3) spin
rotation symmetry is preserved, both the bulk Haldane phase and its
edge state (two free spins) are stable.\cite{Chen2011a,Chen2013} [Any
perturbation preserving the SO(3) spin rotation symmetry can only
affect the $S=1$ object but cannot affect the fractionalized $S=1/2$
free spins.]

It is worth to mention that the two free spins on the boundary can be
revealed from the derivation of the topological term in Eq.(\ref{eq1})
as\cite{Ng1994}
\begin{eqnarray}
  &&iS\sum_{m}(-1)^{m}\Omega_{\text{WZ}}(\vec{n})\simeq i\frac{S}{2}\int_{0}^{L}dx\frac{\partial \Omega_{\text{WZ}}(\vec{n}(x))}{\partial x}\nonumber\\
  &&=i\frac{S}{2}[\Omega_{\text{WZ}}(\vec{n}(x=L))-S_{\text{WZ}}(\vec{n}(x=0))]+i2\pi SQ,\label{eq4}
\end{eqnarray}
where $\Omega_{\text{WZ}}(\vec{n})$ denotes the Berry phase of unit
vector $\vec{n}(m)$ on the $m$-th site. If one does not choose the
periodic boundary condition, the first term will not be canceled and
it is easy to see that they are indeed two free spin-S/2 spins on the
boundary. (A spin-S spin has the Berry phase of
$S\Omega_{\text{WZ}}(\vec{n})$.) In our case, we observe that the effective
spin-1 chain has the non-trivial edge state (two free spins).

In the opposite limit with $V\rightarrow+\infty$ (antiferromagnetic
coupling between two chains), one may set
$\vec{n}_{1}=-\vec{n}_{2}\equiv\vec{n}$ and the resulting action
simply reads
\begin{eqnarray}
S=\int d\tau dx\frac{1}{2cg}(\partial_{\mu}\vec{n})^{2},\label{eq5}
\end{eqnarray}
where we note that the topological term totally vanishes in this
situation. Thus, in contrast to the ferromagnetic case, no non-trivial
edge state like two free spins appears when the coupling between the
two chains is negative (antiferromagnetic). Since the one-dimensional
NLSM without topological term is gapped and described by the strong
coupling $g\rightarrow+\infty$ at fixed point, we may instead use the
Hamiltonian formalism (quantum rotor) to get an intuitive
understanding (setting $c=1$ below):\cite{Sachdev2011}
\begin{eqnarray}
H=\frac{g}{2}\sum_{i}\hat{\vec{L}}^{2}_{i}+\frac{1}{2g}\sum_{i}\hat{\vec{n}}_{i}\cdot\hat{\vec{n}}_{i+1}\label{eq6},
\end{eqnarray}
where we have the commutation relations
$[\hat{n}_{i\alpha},\hat{n}_{j\beta}]=i\delta_{ij}\epsilon_{\alpha\beta\gamma}\hat{n}_{i\gamma}$,
$[\hat{L}_{i\alpha},\hat{L}_{j\beta}]=i\delta_{ij}\epsilon_{\alpha\beta\gamma}\hat{L}_{i\gamma}$
and
$[\hat{L}_{i\alpha},\hat{n}_{j\beta}]=i\delta_{ij}\epsilon_{\alpha\beta\gamma}\hat{n}_{i\gamma}$.
$\hat{\vec{n}}$ acts like the canonical position and $\hat{\vec{L}}$
corresponds to the angular momentum. The eigenstates of
$\hat{\vec{L}}^{2}$ read $|l,m\rangle$ with $-l\leq m\leq l$ and $l,m$
being integers. When $g$ is large, the ground state is the
product-state $|\Psi\rangle\simeq\prod_{i}|l=0\rangle_{i}$. The first
excitation state is a triplet with $|l=1,m=0,\pm1\rangle$ on certain
site, which corresponds to the physical $S=1$ triplet
excitation.\cite{Sachdev2011}

\begin{figure}
\includegraphics[width=0.8\columnwidth]{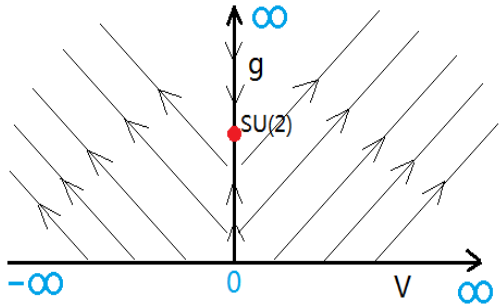}
\caption{\label{fig:2} A qualitative RG flow diagram for the two
  coupled spin-1/2 antiferromagnetic Heisenberg chains. The SU(2)
  point represents the gapless fixed point of the single spin-1/2
  antiferromagnetic chain.}
\end{figure}

Furthermore, when $V$ deviates from the above three limits, the
standard perturbative renormalization group (RG) [see
Fig.~\ref{fig:2}] tells us that even small perturbation around the
fixed point $V=0$ will drive the system to either strong coupling
fixed point $V\rightarrow-\infty$ or
$V\rightarrow+\infty$.\cite{Giamarchi} In contrast, the latter two are
rather stable for small perturbations and can be identified as two
genuine phases. Therefore, we expect the qualitative analysis
presented here may capture the basic physics. (The above features can
also be obtained by standard abelian bosonization on two coupled
spin-1/2 antiferromagnetic Heisenberg chains.\cite{Gogolin,Giamarchi})

\section{The Kondo necklace model}\label{sec3}

Now, it is ready to study the Kondo necklace model, which is a highly
simplified model only including spin degree of freedom for both local
and conduction electrons at half-filling.\cite{Doniach}
\begin{eqnarray}
H_{\text{KN}}=t\sum_{i}(\tau_{i}^{x}\tau_{i+1}^{x}+\tau_{i}^{y}\tau_{i+1}^{y})+J\sum_{i}\vec{\tau_{i}}\cdot\vec{S_{i}}\label{eq7}
\end{eqnarray}
where $\tau_{i}^{x},\tau_{i}^{y}$ represents the spin degrees of
freedom coming from original conduction electrons and $\vec{S}_{i}$
denoting local spins. The first term is like a one-dimensional quantum
XY model and can be easily solved by using the Jordan-Wigner
transformation
$\tau_{i}^{+}=2c_{i}^{\dag}e^{i\pi\sum_{-\infty}^{i-1}c_{j}^{\dag}c_{j}},\tau_{i}^{z}=2c_{i}^{\dag}c_{i}-1$. The
resulting Hamiltonian reads
$H_{XY}=2t\sum_{i}(c_{i}^{\dag}c_{i+1}+c_{i+1}^{\dag}c_{i})$ with
spin-spin correlation
$\langle\tau^{z}(x)\tau^{z}(0)\rangle\sim\frac{1}{x^{2}},\langle\tau^{+}(x)\tau^{-}(0)\rangle\sim\frac{1}{x^{1/2}}$.
The second part of $H_{\text{KN}}$ describes the Kondo coupling
between the local spins and spin degrees of freedom from conduction
electrons, which is fundamental to form the Kondo spin singlet state
(It is also called Kondo spin liquid when the system is insulating).
Based on many analytical and numerical studies, it is well established
that for the 1D Kondo necklace model, its ground-state is
a gapped Kondo spin singlet state for all value of $J/t$ except for
$J=0$.\cite{Scalettar,Moukouri,Nishino,Zhang2000k,Valencia2010a,Valencia2010b}
However, we should remind the reader that the Kondo necklace model
cannot be derived from the original Kondo lattice model at
half-filling but could only be considered as a phenomenological model
devised for studying the low-lying spin excitations.

In the study of the Kondo necklace-like model, an analytical approach
called bond-operator representation could be useful for such spin-only
models.\cite{Sachdev1990,Zhang2000k} When the bond-operator mean-field
approximation is used, the low-lying spin triplet excitation is gapped
for all $J/t>0$, which reproduces the expected results. The existence
of the triplet quasiparticle gap denotes that there only exists a
Kondo spin singlet phase in this model.\cite{Zhang2000k}

\begin{figure}
\includegraphics[width=0.5\columnwidth]{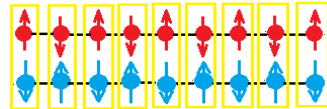}
\caption{\label{fig:3} A cartoon is shown for the Kondo spin liquid
  phase where each pair of conduction and local spins forms a spin
  singlet state. No non-trivial edge state appears in such phase.}
\end{figure}

It is interesting to inspect the low energy properties of the Kondo
necklace model in terms of the effective NLSM. However, since no
exchange interaction exists among any local spins $\vec{S_{i}}$, the
original Kondo necklace model is unable to give rise a useful
formalism of NLSM. To resolve this drawback, we may add the Heisenberg
exchange interaction for local spins. We note that this treatment is
indeed widely used in literatures, which could mimic the effect of
RKKY interaction and is friendly to further approximation
treatments.\cite{Vojta,Senthil2003,Senthil2004} Moreover, one can also
introduce exchange interaction between $z$-component spins
$\tau^{z}$. We argue that the main feature of spin degrees of freedom
of conduction electrons is its gaplessness and such feature does not
rely on the specific XY-like formalism. We can also imagine that there
exists strong interaction among original electrons and this
interaction leads to the exchange interaction of $\tau^{z}$. Then, we
propose the extended Kondo necklace model as following
\begin{eqnarray}
H_{\text{EKN}}=t\sum_{i}\vec{\tau}_{i}\cdot\vec{\tau}_{i+1}+J\sum_{i}\vec{\tau_{i}}\cdot\vec{S_{i}}+W\sum_{i}\vec{S}_{i}\cdot\vec{S}_{i+1},
\end{eqnarray}
and its corresponding NLSM reads as
\begin{eqnarray}
&&S=\sum_{i=1,2}\int d\tau dx\mathcal{L}_{i}+i2\pi SQ_{i}+\int d\tau dx J(\vec{n}_{1}\cdot\vec{n}_{2}),\nonumber\\
&&\mathcal{L}_{i}=\frac{1}{2c_{i}g_{i}}[(\partial_{\tau}\vec{n}_{i})^{2}+c_{i}^{2}(\partial_{x}\vec{n}_{i})^{2}],
\end{eqnarray}
where $c_{1}\sim t$, $c_{2}\sim W$ and $\vec{n}_{1},\vec{n}_{2}$
represent spins of conduction electrons and local moments,
respectively. Clearly, the above NLSM is just the one studied in
previous section (Eq.(\ref{eq1})). But it is important to note $J>0$
in this case. Thus, based on the previous discussion of $J>0$ ($V>0$
in Eq.(\ref{eq1})), we expect the topological term will vanish and the
system can be described by a trivial $S=1$ chain (Eq.(\ref{eq5})), where
the ground-state is a featureless disordered state and no free
spin-1/2 spin exists on either side of the chains with open boundary
condition. Such trivial disordered state should be the expected Kondo
spin liquid state because $\vec{\tau}$ or $\vec{S_{i}}$ chain (both
are effective $S=1/2$ antiferromagnetic chains) alone cannot produce a
gapped state due to their intrinsic $i\pi Q$ topological term. Only
when these two chains interact antiferromagnetically, their $i\pi Q$
topological terms destructively interfere, which gives rise to the
trivial disordered ground-state as shown in Fig. \ref{fig:3}. We also
know that the Kondo singlet state should be formed between
$\vec{\tau}$ and $\vec{S_{i}}$\cite{Tsunetsugu} and in NLSM, it is
described by $\vec{n}_{1}=-\vec{n}_{2}$ in order to form a $S=0$ object
($\vec{S}_{total}=S\vec{n}_{1}+S\vec{n}_{2}=0$). [We recall that
$\vec{n}_{1}=\vec{n}_{2}$ forces these two spins to form a $S=1$ object
($\vec{S}_{total}=S\vec{n}_{1}+S\vec{n}_{2}=1$ with $S=1/2$) as what
have been studied in Eq.(\ref{eq3}) when the Haldane phase is
concerned.] Therefore, we may conclude that the Kondo spin liquid
state is described by the featureless disordered state without
noticeable non-trivial topological properties. In this respect, we may
regard the Kondo spin liquid state as a classical disordered phase due
to the vanished topological terms, which in fact represents the
fundamental quantum features.

Comparing the Kodno spin liquid phase with the $S=1$ Haldane phase, we
may see that their bulk properties seem to be similar since the bulk
excitation are both described by NLSM at strong coupling fixed point
without the contribution form the topological term in closed
manifold. The distinction mainly relies on their boundary, namely
the Haldane phase has fractionalized $S=1/2$ spins while the Kodno spin
liquid does not support such degree of freedom on either side of the
open chains.

\section{Kondo spin liquid in higher spatial dimension and possible
  relation to SPT states} \label{sec4}

Now, if we extend 1D Kondo necklace model to higher space dimension,
besides the Kondo spin liquid phase, the usual antiferromagnetic
insulating state appears (We only consider bipartite lattice as square
and honeycomb
lattices).\cite{Zhang2000k,Langari2006,Reyes2007,Langari2009} It has
been suggested that there may exist a second quantum phase transition
between these two.\cite{Zhang2000k, Reyes2007} Based on previous
discussion, we may considered the 2D Kondo necklace model as a
bilayer antiferromagnet with antiferromagnetic inter-layer
coupling. Each layer alone may form the antiferromagnetic N\'eel state
or valence-bond solid (VBS) phase, while the inter-layer coupling can
drive the system into the Kondo spin liquid phase. In the language of
NLSM, we now have an effective two-dimensional O(3) NLSM but its usual
topological term (Hopf term) vanishes in the strong coupled Kondo
phase, similar to the 1D case.\cite{Fradkin2013} Therefore,
we may conclude that the Kondo spin liquid phase are topologically
trivial and cannot show any non-trivial edge (surface) states if no
other novel elements are included.

However, the above argument is not the whole story. It has been
realized that there are non-smooth solutions of the classical
Euclidean equations of motion of (2+1)-D O(3) NLSM, known as
hedgehogs, which have non-trivial topological (winding)
numbers.\cite{Fradkin2013} Furthermore, if we introduce the VBS order
parameter explicitly into the NLSM, the resulting model would be an
O(4) or O(5) NSLM with appropriate topological terms. Although
the inter-layer Kondo coupling might introduce possible O(3)$\times
Z_{2}$ or O(3)$\times$O(2) anisotropies, since the original
conduction electron do not show spin-rotation or lattice-translation
symmetry-breaking, we may neglect such anisotropies when the Kondo
spin liquid phase is concerned. In the following subsections, we
will focus on the O(4) and O(5) NSLM .

\subsection{A possible topological term in O(4) NLSM and relation to
  2D SU(2) SPT states}

First, we consider the competition between the $(\pi,\pi)$ N\'eel
order and the $(\pi,0)$ VBS order. In this case, the VBS order
parameter is a one-component real scalar
$\phi_{0}$.\cite{Senthil2006} The nontrivial winding number $Q_{H}$ is
expressed in terms of a four-component unit vector
$\hat{\phi}=(\phi_{0},\vec{n})$,
\begin{eqnarray}
Q_{H}=\frac{1}{12\pi^{2}}\int d\tau d^{2}x \epsilon_{\alpha\beta\gamma\delta}\epsilon_{\mu\nu\lambda} \phi_{\alpha}\partial_{\mu}\phi_{\beta}\partial_{\nu}\phi_{\gamma}\partial_{\lambda}\phi_{\delta},
\end{eqnarray}
where $\phi_{0}$, $\vec{n}$ denote the $(\pi,0)$ VBS (dimer) order and
the usual N\'eel order vector,
respectively.\cite{Senthil2006,Xu2011}[However, in the square lattice,
the more general VBS order parameter is complex, which is used to
describe both the columnar and the plaquette ordering
patterns.\cite{Senthil2006} This situation will be analyzed in next
subsection as well.]

The resulting NLSM for two coupled antiferromagnetic layers with
antiferromagnetic inter-layer coupling reads
\begin{eqnarray}
&&S=\sum_{i=1,2}\int d\tau d^{2}x\mathcal{L}_{i}+i\pi Q_{i}+\int d\tau d^{2}x V(\vec{n}_{1}\cdot\vec{n}_{2}),\nonumber\\
&&\mathcal{L}_{i}=\frac{1}{2c_{i}g_{i}}(\partial_{\mu}\hat{\phi}_{i})^{2},\nonumber\\
&&Q_{Hi}=\frac{1}{12\pi^{2}}\int d\tau d^{2}x \epsilon_{\alpha\beta\gamma\delta}\epsilon_{\mu\nu\lambda} \phi_{i\alpha}\partial_{\mu}\phi_{i\beta}\partial_{\nu}\phi_{i\gamma}\partial_{\lambda}\phi_{i\delta}.\label{eq8}
\end{eqnarray}
Due to the $V$-term which reflects the Kondo coupling between
conduction and local spin degree of freedom, the above NLSM is not
O(4) invariant but degenerates into global O(3)$\times Z_{2}$
symmetry. The $Z_{2}$ refers to the VBS order parameter $\phi_{0}$,
while the O(3) invariant corresponds to the rotation symmetry of
$\vec{n}_{1,2}$. Here, unfortunately, we do not know how to treat even
the strong coupling limit $V\rightarrow\pm\infty$. Instead, in order
to restore the O(4) invariant, we may artificially add the term
$V\phi_{01}\phi_{02}$. [It seems that if certain multi-spin exchange
interaction is added, this O(4) symmetry may be realized
microscopically, as indicated from the example of four-spin
interaction in the two coupled one-dimensional chains.\cite{Gogolin}]
Then, interestingly, one finds that the strong coupling limit
$V\rightarrow\pm\infty$ is described by the same action (setting
$c=1$)
\begin{eqnarray}
S=\int d\tau d^{2}x\frac{1}{g}(\partial_{\mu}\hat{\phi})^{2}+i2\pi Q_{H}.\label{eq9}
\end{eqnarray}
When $g$ is large, this model disorders (the Kondo spin liquid phase)
and its bulk excitation is gapped. The $i2\pi Q_H$ term contributes a
factor of unity to the partition function and has no noticeable
contribution for the bulk spectrum. Interestingly, the boundary of
this O(4) model is gapless because the edge state is described by the
(1+1)-D $SU(2)_{k=1}$ WZW model.\cite{Xu2011} This point can be made
clear as follows. First, using the bulk-boundary correspondence, the
edge action reads\cite{Xu2011}
\begin{eqnarray}
&&S=\int d\tau dx\frac{1}{g}(\partial_{\mu}\hat{\phi})^{2}\nonumber\\
&&+i\frac{2\pi}{12\pi^{2}}\int d\tau dx\int_{0}^{1}du \epsilon_{\alpha\beta\gamma\delta}\epsilon_{\mu\nu\lambda} \phi_{\alpha}\partial_{\mu}\phi_{\beta}\partial_{\nu}\phi_{\gamma}\partial_{\lambda}\phi_{\delta}.\label{eq10}
\end{eqnarray}
Here the second term is usually called the Wess-Zumino term. The field
$\hat{\phi}(x,\tau,u)$ is defined on a three-dimensional hemisphere
with a boundary coinciding with the two-dimensional plane $(x,\tau)$,
where the original theory is defined. So we have
$\hat{\phi}(x,\tau,u=0)=\hat{\phi}(x,\tau)$.\cite{Gogolin} Being a
total derivative, the three-dimensional integral does not really
depend on $\hat{\phi}(x,\tau,u)$.\cite{Gogolin} By introducing an
SU(2) matrix
$\hat{U}=\phi_{0}\hat{I}+i\vec{n}\cdot\hat{\vec{\sigma}}$,\cite{Senthil2006,Xu2011}
the above action can be rewritten as the standard SU(2)$_{k=1}$ WZW
model
\begin{eqnarray}
&&S=\int d\tau dx\frac{1}{2g}Tr(\partial_{\mu}\hat{U}^{-1}\partial_{\mu}\hat{U})+\nonumber\\
&&\frac{2\pi i}{24\pi^{2}}\int d\tau dx du \epsilon_{\mu\nu\lambda}Tr (\hat{U}^{-1}\partial_{\mu}\hat{U}\hat{U}^{-1}\partial_{\nu}\hat{U}\hat{U}^{-1}\partial_{\lambda}\hat{U}).\label{eq11}
\end{eqnarray}
And the ground-state is described by the following gapless fixed
action (also called critical WZW model\cite{Gogolin})
\begin{eqnarray}
&&S=\int d\tau dx\frac{1}{8\pi}Tr(\hat{U}^{-1}\partial_{\mu}\hat{U}\hat{U}^{-1}\partial_{\mu}\hat{U})+\nonumber\\
&&\frac{i}{12\pi}\int d\tau dxdu \epsilon_{\mu\nu\lambda}Tr (\hat{U}^{-1}\partial_{\mu}\hat{U}\hat{U}^{-1}\partial_{\nu}\hat{U}\hat{U}^{-1}\partial_{\lambda}\hat{U}).\label{eq12}
\end{eqnarray}
For the WZW model Eq.(\ref{eq11}) and Eq.(\ref{eq12}), the symmetry is $SU(2)_{L}\times SU(2)_{R}$,
which states the action is invariant under transformation $\hat{U}\rightarrow G_{L}\hat{U}G_{R}$ with
$G_{L},G_{R}$ being SU(2) matrices. Such $SU(2)_{L}\times SU(2)_{R}$ symmetry is in fact related to two decoupled
edge excitation for the critical WZW model Eq.(\ref{eq12}). One is the left mover $J_{+}=\frac{K}{2\pi}\partial_{+}\hat{U}\hat{U}^{-1}$ and
other is $J_{-}=\frac{-K}{2\pi}\hat{U}^{-1}\partial_{-}\hat{U}$. (We have defined the light-cone or chiral coordinate as $x^{\pm}=\frac{1}{\sqrt{2}}(\tau\pm ix)$
and $\partial_{\pm}=\frac{1}{\sqrt{2}}(\partial_{\tau}\mp i\partial_{x})$.) Those two movers satisfy the equation of motion $\partial_{\mp}J_{\pm}=0$ and $J_{-}$ is
invariant under $SU(2)_{L}$ transformation ($\hat{U}\rightarrow G_{L}\hat{U}$) while $J_{+}$ transforms as $G_{L}J_{+}G_{L}^{-1}$. Thus, only the left mover $J_{+}$
carries the $SU(2)_{L}$ charge and $J_{-}$ is $SU(2)_{L}$ neutral. Furthermore, if a suitable SU(2) external field is introduced, the left mover $J_{+}$ will response to it and
one may see a quantized Hall conductance.\cite{Liu2013}

A careful reader may notice that the Eq.(\ref{eq9}) is just the action
of a kind of (2+1)-D SU(2) SPT states, and Eq.(\ref{eq12}) corresponds
to the expected symmetry-protected gapless edge state of the SU(2) SPT
state.\cite{Liu2013} However, we should emphasize that the original
model Eq.(\ref{eq8}) does not have the O(4) invariant due to the
$V$-term.  In other words, the SU(2) symmetry (in fact the $SU(2)_{L}$), which leads to the
symmetry-protected gapless edge state, will not be
preserved. Therefore the disorder phase of our original model
(Eq.(\ref{eq8})) is not an SPT state but seems a trivial one. However,
we hope that the present study may help to identify certain
microscopic bilayer antiferromagnets models which can support the
desirable SPT state.

\subsection{O(5) NLSM and relation to 3d SPT states}

It is noted that in the square lattice, the more general VBS order
parameter should be a complex field, in order to describe both the
columnar and the plaquette ordering patterns.\cite{Senthil2006} In
this case, the complex VBS order parameter reads as
$\psi_{\text{VBS}}=\phi_{vx}+i\phi_{vy}$, with $\phi_{vx},\phi_{vy}$
being real fields. An O(5) vector field
$\hat{\phi}=(\phi_{vx},\phi_{vy},\vec{n})$ can be constructed
accordingly, where the later three-component vector field $\vec{n}$ is
the usual N\'eel order parameter. Therefore, the effective O(4) NSLM
(Eq.(\ref{eq9})) considered in the previous subsection has to be
replaced by an O(5) version of NLSM.\cite{Senthil2006}
\begin{eqnarray}
S=\int d\tau d^{2}x\frac{1}{g}(\partial_{\mu}\hat{\phi})^{2}+i2\pi \widetilde{Q},
\end{eqnarray}
where the winding number $\widetilde{Q}$ is defined as
\begin{eqnarray}
\widetilde{Q}=\frac{1}{64\pi^{2}}\int du d\tau d^{2}x \epsilon_{\alpha\beta\gamma\delta\rho}\epsilon_{\mu\nu\lambda\vartheta} \phi_{\alpha}\partial_{\mu}\phi_{\beta}\partial_{\nu}\phi_{\gamma}\partial_{\lambda}\phi_{\delta}\partial_{\vartheta}\phi_{\rho}.\nonumber
\end{eqnarray}
Here, the $i2\pi \widetilde{Q}$ is also known as the WZW term, and
$\hat{\phi}(\tau,x,y,u)$ is an extension of the space-time
configuration of $\hat{\phi}(\tau,x,y)$, which satisfies
$\hat{\phi}(\tau,x,y,0)=(0,0,0,0,1)$ and
$\hat{\phi}(\tau,x,y,1)=\hat{\phi}(\tau,x,y)$. Interestingly, such
(2+1)-D O(5) NLSM describes one of the gapless surface states of 3D SPT
phase, whose effective action is\cite{Xu2013}
\begin{eqnarray}
&&S=\int d\tau d^{3}x\frac{1}{g}(\partial_{\mu}\hat{\phi})^{2}\nonumber\\
&&+i\frac{2\pi}{64\pi^{2}}\int d\tau d^{3}x \epsilon_{\alpha\beta\gamma\delta\rho}\epsilon_{\mu\nu\lambda\vartheta} \phi_{\alpha}\partial_{\mu}\phi_{\beta}\partial_{\nu}\phi_{\gamma}\partial_{\lambda}\phi_{\delta}\partial_{\vartheta}\phi_{\rho}.\nonumber
\end{eqnarray}

Since it has been known that the O(5) NLSM with the WZW term is
gapless, it seems that it can not describe the gaped Kondo spin liquid
phase. On the other hand, if the WZW term is irrelevant in
this situation, we might suggest that an O(5) NLSM without the WZW
term may reproduce the correct physics when the Kondo spin liquid
state is concerned. Again, it is consistent with the previous
speculation that the state belongs to a trivial disorder phase.

\subsection{An alternative route to the description of
  high-dimensional Kondo spin liquid}

An alternative route to describe the Kondo necklace model in 2D is by
using (coupled multi-chains) multi-leg ladders for each effective
antiferromagnetic layer.\cite{Gogolin} However, the multi-leg-ladder
construction itself suffers from the serious even-odd effect, where
the physics heavily depends on whether the number of legs (chains) is
even or odd.\cite{Gogolin} For a odd-leg ladder, it behaves like the
single spin-1/2 Heisenberg chain, while an even-leg ladder has similar
properties as the $S=1$ Heisenberg chain.\cite{Gogolin} Based on the
consideration that the realistic 2D system should not have such strong
even-odd effect, we may not use this formalism in our discussion. But
it is interesting to point out that there exist some interesting
studies of multi-leg spin ladders. For example, see
Ref.[\onlinecite{Capponi2013}] and references therein.

\subsection{Comparison with topological Kondo insulator}

In addition, we note that in literature, the so-called topological
Kondo insulator has been proposed as an extension of the original
topological (band) insulators.\cite{Dzero} In contrast to usual topological
insulators where the spin-orbit coupling is encoded in a
spin-dependent hopping amplitudes between different unit cells, the
topological Kondo insulator is produced by the spin-orbit coupling
associated with the hybridization between conduction and local
electrons.\cite{Dzero} Different from our case, these
insulators are obviously topological non-trivial due to their
intrinsic spin-orbit couplings, and their gapless surface states are
protected by the topology of the bulk (time-reversal symmetry).

\section{Conclusion} \label{sec5}

In summary, we have studied the Kondo necklace model in terms of
effective non-liner sigma model field theory and found that the bulk
properties of the Kodno spin liquid phase are similar to the
celebrated Haldane. The distinction of these two phases mainly relies
on their boundary, where on either side of the open chains, the
Haldane phase has a fractionalized $S=1/2$ spin while the Kondo spin
liquid does not support such degree of freedom. In addition, we
analyze the topological properties of Kondo spin liquids in high
dimensional Kondo necklace models. It seems that those Kondo spin
liquids are still topologically trivial compared to the topological
Kondo insulator unless some spin-orbit coupling-like elements are
introduced. Yet, interestingly, if the system's symmetry is extended
to O(4), and the hedgehog-like point defects are taken into account,
our model can be identified to a kind of SU(2) SPT models. This
indicates that certain microscopic bilayer antiferromagnets models
could support the desirable SPT state and even some real-world bilayer
materials might be devised accordingly in the near
future. Furthermore, if the symmetry is expanded to O(5), the
effective action just describes the surface state of a 3D SPT state
though such O(5) NLSM could not be a proper description of Kondo spin
liquid phase due to its gaplessness. We expect the present work may be
useful for further study on states with nontrivial topological
properties.

\begin{acknowledgments}

  We thank Cenke Xu for useful discussion on the O(5) NLSM. The work
  was supported partly by NSFC, PCSIRT (Grant No. IRT1251), the
  Program for NCET, the Fundamental Research Funds for the Central
  Universities and the national program for basic research of China.

\end{acknowledgments}

\end{document}